\documentstyle[11pt,moriond,psfig]{article}
\def\be{\begin{equation}}
\def\ee{\end{equation}}
\def\bea{\begin{eqnarray}}
\def\eea{\end{eqnarray}}

\begin{document}
\vspace*{2cm}
\title{COLOR DEGREES OF FREEDOM IN NUCLEAR COLLISIONS AND CHARMONIUM SUPPRESSION}

\author{J\"ORG H\"UFNER$^{a,b}$, BORIS Z. KOPELIOVICH$^{b,c}$ 
	and ALBERTO POLLERI$^{a}$}

\address{$^{a}$Institut f\"ur Theoretische Physik der Universit\"at, 
	Philosophenweg 19, 69120 Heidelberg, Germany. \\[0.5mm]
	$^{b}$Max Planck Institut f\"ur Kernphysik, Postfach 103980, 
        69029 Heidelberg, Germany. \\[0.5mm]
	$^{c}$Joint Institute for Nuclear Research, Dubna, 
	141980 Moscow Region, Russia. \\[2mm]}

\maketitle\abstracts{
In high energy nuclear collisions, the Glauber model is commonly used to
evaluate $J/\psi$ suppression due to inelastic 
interaction with colorless bound nucleons. This requires an effective value for the
$J/\psi$-nucleon absorption cross section which is larger than theoretically
expected. On the other hand, multiple nucleon-nucleon collisions mediated by color
exchange interactions, excite their color degrees of freedom. We investigate the 
importance of this effect and conclude that the related corrections are 
important to explain the effective value extrapolated from experiment.}

In a nucleus-nucleus ($AB$) collision $J/\psi$ suppression is mostly treated 
within the Glauber model, with a constant effective cross section for 
absorption on bound, non-interacting nucleons. Here we intend to study and to improve
the understanding of the physical phenomena behind early stage absorption.
Following a known interpretation of the dynamics of nucleon-nucleon ($NN$) 
collisions, we ascribe the main contribution to the inelastic cross section to 
color exchange processes. Via color exchange,
the 3-quark systems constituting the colliding nucleons become color octet states.
Color exchange may also be accompanied by gluon radiation. While
the contribution of radiated gluons to $J/\psi$ absorption was treated 
in \cite{HK98HuHK00}, here we focus on the first part. Having realized that
color-singlet nucleons are found in color octet states after an elementary 
collision, one can expect that repeated scatterings lead to multiple color exchanges. 
One therefore has the possibility to excite the color degrees of freedom of the
3-quark system also to the decuplet state. Higher color representations can be 
obtained when radiated gluons are also present, but here we restrict our study on
the color dynamics of the constituent quarks of a nucleon.

To calculate the value of the inelastic cross section for $\Psi$ on a nucleon
which has undergone multiple scattering, we use the two-gluon exchange model of 
Low and Nussinov \cite{LN75GS77}. Although oversimplified, the model
incorporates important properties of high-energy hadronic cross sections such as 
(approximate) energy independence and scaling of the cross sections according to
the mean square radius of hadrons. Indicating with $\Psi$ the effective charmonium 
state, composed by a mixture of $J/\psi$, $\psi'$ and $\chi$,
one can write at once the absorption cross section, obtaining
\be
\sigma^{abs}_{\Psi N_n} = \int d^2\vec{k} \ \frac{1}{(k^2 + \mu^2)^2}
\ \mbox{{\large $\langle$} $\!\!\Psi\!\!$ {\large $|$}}
\, 1 - \exp(\,i\, \vec{k}\,\vec{x}_{12}\,)
\, \mbox{{\large $|$} $\!\!\Psi\!\!$ {\large $\rangle$}} 
\ \mbox{{\large $\langle$} $\!\!N_n\!\!$ {\large $|$}}
\,\Gamma^a(k) \,\Gamma^{\dagger\,a}(k)
\,\mbox{{\large $|$} $\!\!N_n\!\!$ {\large $\rangle$}} \,.
\label{colorcross}
\ee
The first factor in the integrand is the squared gluon propagator,
the second contains the phase shift due to the momentum transfer $\vec k$ and is 
responsible, by the way, for color transparency,
while the last factor is the interesting part of the present calculation. It is 
the product of two vertices
between three quarks and a gluon, averaged with the wave function of the 
colored nucleon. The vertex is
$\Gamma^a(k) = 16/3 \, \alpha_s \,[
\, \lambda^a_{(1)})\, \exp(\,i\, \vec{k}\,\vec{y}_1\,)
+ \lambda^a_{(2)}\exp(\,i\, \vec{k}\,\vec{y}_2\,) +
\lambda^a_{(3)} \exp(\,i\, \vec{k}\,\vec{y}_3\,)]$.
In order to average the product of the two vertices, one needs to specify the
structure of the nucleon wave function. The color part of the wave 
function is the relevant one, while the spatial part is chosen
to be equal for all color multiplets. The factor given by the angular brackets 
in eq.$\,$(\ref{colorcross}) can be evaluated and leads to
\be
\mbox{{\large $\langle$} $\!\!N_n\!\!$ {\large $|$}}
\Gamma^a\,\Gamma^{\dagger\,a} 
\mbox{{\large $|$} $\!\!N_n\!\!$ {\large $\rangle$}}
= \mbox{{\large $\langle$} $\!\!N\!\!$ {\large $|$}}
\sum_{i < j}\, 1 - c_n \exp(\,i\, \vec{k}\,\vec{y}_{\,ij}\,)
\mbox{{\large $|$} $\!\!N\!\!$ {\large $\rangle$}}.
\label{colourcrossfin}
\ee
The effect of having different color states is all contained in the
coefficients $c_n$, listed in the second column of Tab.~\ref{psinuc}. 
The values of $c_n$ found, for $n = {\bf 8},\,{\bf 10}$, lead to larger absorption
cross sections as compared to the singlet case.
To see this it is useful to define the two-quark 
form factor $F^{2q}_N(k^2) = \mbox{{\large $\langle$} $\!\!N\!\!$ {\large $|$}}
\exp(\,i\,\vec{k}\,\vec{x}_{\,ij}\,)
\mbox{{\large $|$} $\!\!N\!\!$ {\large $\rangle$}}$
for the nucleon. Since, within our approximation, the latter
does not depend on the colour multiplet, we drop the label $n$. Then we can
rewrite eq.$\,$(\ref{colorcross}) as $ \sigma^{abs}_{\Psi N_n} = 
\mbox{{\large $\langle$} $\!\!\Psi\!\!$ {\large $|$}} \,\sigma_n(r)\,
\mbox{{\large $|$} $\!\!\Psi\!\!$ {\large $\rangle$}}$,
where 
\be
\sigma_n(r)  =  \mbox{\large $\frac{16}{3}$} \ \alpha_s^2 
\int d^2\vec{k} \ \frac{1}{(k^2 + \mu^2)^2}
\  [ 1 - \exp(\,i\, \vec{k}\,\vec{r}\,)]\ [1 - c_n F^{2q}_N(k^2)]\,,
\label{dipole}
\ee
exhibiting a similar structure to that of the usual dipole-nucleon cross section, 
but with the important difference carried by the coefficients $c_n$.
\begin{table}[b]
\caption{Results of the calculation of the meson-colored nucleon cross sections.}
\begin{center}
\begin{tabular}{|c||c|c|c|} \hline
Multiplet $\ $ \{n\} & $\ \ c_n\ \ $ & $\ \sigma_{\psi N_n}$ [mb] $\ $ 
	& $\ \ \delta_n \ \ $\\ \hline\hline
Singlet  $\ \ \ \;$ \{{\bf 1}\}   &   $\ 1$    & 5.8 &  1.0   \\ \hline
Octet  $\ \ \ \ \ $ \{{\bf 8}\}   & $\, -1/4\ $ & 7.8 &  1.35   \\ \hline
$\ $ Decuplet $\ $\{{\bf 10}\} $\ $ &  $\ \,1/2$   & 9.9 &  1.7   \\ \hline
\end{tabular}
\end{center}
\label{psinuc}
\end{table}
Notice that with $c_n \neq 1$ and $\mu = 0$ the integral in eq.$\,$(\ref{dipole})
is infrared divergent. Therefore, the cut-off $\mu$, of the order of the inverse
confinement radius, is important. With $\mu \neq 0$ it holds that
 $\sigma_n(r) \rightarrow 0$ for $r \rightarrow 0$, so 
that one still preserves color transparency, even with a colored nucleon. This is
indeed a property of the singlet meson. Here we are not primarily interested
in the absolute values of $\sigma^{abs}_{\Psi N_n}$ but rather on its dependence
on the color state $n$ with respect to the singlet value. Yet we estimated the 
absolute values choosing $\alpha_s = 0.6$ and $\mu = 140$ MeV. Within a harmonic
oscillator model of the nucleon, one finds that $F^{2q}_N(k^2) = F^{em}_N(3k^2)$,
the electromagnetic form factor. We assume that this property is general and
take the dipole form $F^{em}_N(Q^2) \sim (\lambda^2 + Q^2)^{-2}$
with the parameter $\lambda^2 = 12 / R^2_p$ and $R_p = 0.8$ fm.
With the above definitions, the generalized dipole-nucleon cross section can be 
computed analytically
and the results of the calculation are illustrated in Fig.~\ref{FIGDIP}, 
showing the modified dipole-nucleon cross section for the three color
states of the nucleon. It largely increases when the dipole scatters off a
colored object. One can get more insight in the effect of the different color
states rewriting eq.$\,$(\ref{dipole}) in the form
\be
\sigma_n(r) = \sigma(r) \left [\, 1 + (1 - c_n)\,\Delta(r) \,\right]\,,
\label{deltadip}
\ee
where $\sigma(r)$ is the usual dipole cross section on a color singlet nucleon
and the function 
$\Delta(r)$ is found to vary appreciably with $r$. Its functional dependence is
given in the small box in Fig. \ref{FIGDIP}. To calculate the inelastic cross section,
we must average $\sigma_n$ with the meson wave function. Therefore, the dependence
of $\Delta$ on $r$ shows that the effect of having colored nucleons is stronger for
larger size mesons.  Recent calculations \cite{HIKT}, based on realistic phenomenology
for the dipole cross section and charmonium wave function, gave values for
cross sections of various charmonia on a nucleon. At center of mass energy $\sqrt{s}
= 10$ GeV, it was found that $\sigma_{J/\psi N} = 3.6$ mb, $\sigma_{\psi' N} = 
12.2$ mb and
$\sigma_{\chi N} = 9.1$ mb. With a composition of $\Psi$ of 52-60$\%$ due to 
$J/\psi$, 8-10$\%$ due to $\psi'$ and 32-40$\%$ due to $\chi$, a weighted value 
of 5.8 mb was obtained. 
In the present case we evaluate $\sigma^{abs}_{\Psi N_n}$ with
a Gaussian wave function having root mean square transverse separation 
$\langle r^2 \rangle^{1/2} = 
0.25$ fm, in order to account for the different states 
implicitly included in $\Psi$, and therefore adjust the singlet cross section to the 
quoted 5.8 mb.
The values found are given in the third column of Tab.~\ref{psinuc}. In order to
minimize the model dependence, we factored out from the colored cross sections the
amount corresponding to the singlet case, therefore writing
$\sigma^{abs}_{\psi N_n} = \sigma^{abs}_{\psi N} \ \delta_n$.
The values for $\delta_n$ are given in the last column of Tab.~\ref{psinuc}.
They constitute the important result of this calculation ,
showing that $\Psi$ scatters with an color octet nucleon with a $35\,\%$ larger
cross section with respect to a singlet nucleon, while with the decuplet the
increase is even of $70\,\%$. The large values found now require a careful analysis
to establish the validity of this newly proposed absorption mechanism.

Since the absorption pattern observed in $pA$ collisions is usually 
considered to be the baseline for the $AB$ case, it is necessary to 
examine it in some detail. 
Before we analyze the data for charmonium production in $pA$ collisions, we
emphasize the importance of charmonium formation time \cite{KZ91HK96}. Within a 
simplified version of a two-channel model \cite{HeHK},
the $\Psi N$ absorption cross section evolves in the eigentime $\tau$ of $\Psi$ as
$\sigma^{abs}_{\Psi N}(\tau) = \sigma^{in}_{\Psi N} + (\sigma^{(0)}
- \sigma^{in}_{\Psi N})\,\cos(\tau/\tau_f)$,
where $\tau_f = (m_{\psi'} - m_{J\psi})^{-1} = 0.3$ fm is the charmonium formation
time, $\sigma^{in}_{\Psi N}$ is the inelastic $\Psi N$ cross section and
$\sigma^{(0)}$ is the cross section 
of the so called ``pre-meson'' on a 
nucleon. As mentioned in the last section, calculations give 
$\sigma^{in}_{\Psi N} = 5.8$ mb \cite{HIKT}. Unfortunately $\sigma^{(0)}$ has
not yet been calculated. We therefore use the value 2.7 mb from \cite{HeHK}. 

One can then write the expression for $\Psi$ production in $pA$ collisions, as
\be
\sigma_\Psi^{pA} = \sigma_\Psi^{NN} / A 
\,\int d^2\vec{b}\,dz \times\,\left[ 1 - \int_z^{+\infty}\!\!\!dz'
\ \sigma^{abs}_{\Psi N}\mbox{\large$($}\mbox{\large $\frac{z'-z}{\gamma_\Psi}$}
\mbox{\large$)$}\ \rho_A(z,\vec{b})/A \right]^{A-1}\label{glaufirst} 
\ee
We want to compare this expression with the conventional one obtained in the
Glauber model, containing
only an effective absorption cross section, therefore making the replacement
$\sigma^{abs}_{\Psi N}(\tau) \rightarrow \sigma^{eff}_{\Psi N}$.
For $A = 208$ and $p_\Psi = $ 50 GeV, using a realistic profile for Pb we obtain 
$\sigma^{eff}_{\Psi N} = 3.8$ mb. This number has to be interpreted as the value 
of the effective absorption cross section of $\Psi$ on a color singlet nucleon,
including formation time effects. To obtain the corresponding value for the case
when $\Psi$ scatters on a color octet nucleon we have to enhance the singlet
result by 35$\%$, obtaining 5.1 mb.

We have then re-analyzed the existing data taken by 3 fixed target 
experiments: NA3 at CERN \cite{NA3}, NA38/51 \cite{NA38/51} at CERN and E866 at 
FERMILAB \cite{E866}. In order remove the effects on absorption
due to different $\Psi$ momenta, we impose the same kinematic conditions, 
considering $\Psi$'s of 50 GeV momentum in the laboratory frame
for the three experiments. We then assume that the 
absorption mechanism is parameterized by the effective cross section 
$\sigma^{eff}_{\Psi N}$. We made three separate fits including
quoted statistical and systematic errors and
extracted the effective absorption cross section by minimizing $\chi^2$.
The result of these fits is presented in the second column of Tab.~\ref{pafit}. 
\begin{table}[b]
\caption{Results of the fits to pA data for the various experiments considered.
See text for details.}
\begin{center}
\begin{tabular}{|c||c|c|} \hline
Experiments & \multicolumn{2}{c|}{$\sigma^{abs}_{\psi N}$ [mb]} \\\cline{2-3}
$p_\psi$ = 50 GeV & Glauber fit & Gluon corr.   \\ \hline\hline
    NA3         &$\ $ 4.5 $\pm$ 1.7 $\ $&$\ $ 7.3 $\pm$ 1.9 $\ $ \\ \hline
    NA38/51     &$\ $ 7.1 $\pm$ 1.6 $\ $&$\ $ 10 $\pm$ 1.8 $\ $ \\ \hline
    E866        &$\ $ 2.7 $\pm$ 0.7 $\ $&$\ $ 4.7 $\pm$ 0.7 $\ $ \\ \hline\hline
    Averages    &$\ $ 3.5 $\pm$ 0.6 $\ $&$\ $ 5.7 $\pm$ 0.6 $\ $ \\ \hline\hline
 Theory & \multicolumn{2}{c|}{3.8 \{{\bf 1}\} $\ \ \ \ $ 5.1 \{{\bf 8}\}} \\ \hline
\end{tabular}
\end{center}
\label{pafit}
\end{table}
The mean value of $3.5 \pm 0.6$ mb seems to favor the singlet mechanism, for which a 
cross section of 3.8 mb is theoretically expected. 

On the other hand, the assumption that $\Psi$ is produced on a bound nucleon with
the same cross section as on a free one, can be an oversimplification of the
physics under study.
In a nuclear environment parton distributions are substantially 
modified. In particular, the gluon distribution in nuclei
exhibits anti-shadowing effects at $x_{Bj} \simeq 0.1$.
We expect that charmonium production on a nucleus is enhanced by a 
factor $R^g_A(x_2,Q^2) = g_A(x_2,Q^2)/g(x_2,Q^2)$. We estimate these effects
by rescaling the $\Psi$ cross section with the A-dependent correction 
$R_A^g = 0.032 \log(A) + 1.006$, extracted from \cite{EKS99}. We fitted the data
once more, obtaining the results listed in the third column of Tab.~\ref{pafit}.
One sees that the new average value of $5.7 \pm 0.6$ mb now seems to favor the
octet mechanism, for which we estimated a value of 5.1 mb. 
One sees that the situation is, in all, quite uncertain. 

The above analysis assumes that there is no dependence of nuclear
effects on the proton energy if the charmonium energy is the same in all
experiments. However, there might still be some dependence on the proton energy.
According to \cite{KN84J00}, energy loss of projectile partons in the nuclear medium
substantially reduces the production rate of charmonia at large $x_F$ and is still
rather important at small $x_F$. In order to compensate for energy
loss, the gluon distribution must be evaluated at a higher value of $x_1$, therefore
suppressing $\Psi$ production. This will in turn reduce the final state absorption on
nucleons in a fit to the data. To account for this effect, further study is required.

To summarize, the data seem to suggest that $\Psi$ is absorbed by colored nucleons
in pA collisions. The picture is though not yet conclusive due to the large 
experimental errors. Nevertheless, we follow our theoretical prediction and examine
the potential role of the color mechanism in the more complex case of $AB$
collisions. Preparing the ground for the forthcoming discussion, it is useful to 
recall the conventional expression within the Glauber model for the $\Psi$ production 
cross section at impact parameter $\vec{b}$. One has
\bea
\frac{d^2\sigma_\Psi^{A B}}{d^2\vec{b}}(\vec{b}) & = & 
\!\!\int d^2\vec{s}\ dz_A\,dz_B \ \sigma_\Psi^{NN}
\ \rho_A(z_A,\vec{s})\ \rho_B(z_B,\vec{b}-\vec{s})\nonumber \\
& \times & \left[1 - \sigma^{abs}_{\Psi N}\,T_A(z_A,\vec{s})/A \right]^{A-1}
\left[1 - \sigma^{abs}_{\Psi N}\,
T_B(z_B,\vec{b}-\vec{s})/B\right]^{B-1}\!\!.
\label{glauber}
\eea
Here we neglect the effects of any additional suppression factors which go beyond
conventional nuclear effects, that is final state interactions with the produced 
matter (QGP, hadron gas, etc.), since we are not primarily interested in these 
contributions. On the other hand, we focus on the reinterpretation of the
terms describing nuclear effects discussing
color exchange in the context of multiple scattering, then applying
the obtained results to the calculation of $\Psi$ production in $AB$ collisions.

Having completed the computation of the $\Psi$ cross sections with colored nucleons,
we need to look at how these color objects develop in a collision between nuclei.
Consider a nucleon of one of the nuclei colliding with a row of nucleons of the
other nucleus. It undergoes multiple scattering, with repeated color exchanges.
It is possible to calculate the probability
$P_n(z,\vec{b})$ for a nucleon in color state 
$n = \,${\bf 1},{\bf 8},{\bf 10} after having traveled through the row up to $z$ 
at impact parameter $\vec{b}$. This can be done in a very elegant way, by solving an
evolution equation for the color density matrix of the 3-quark system \cite{KL83}. 
One obtains
\be
P_n(z,\vec{b}) \!=\! \frac{1}{27} \times\!
\left\{
\begin{tabular}{cc} 
$\ 1 + 20\ F(z,\vec{b}) +\ 6\ G(z,\vec{b})$    &  $\ \{\mbox{\bf 1}\} $   \\ 
$ 16 - 40\ F(z,\vec{b}) + 24\ G(z,\vec{b})$    &  $\ \{\mbox{\bf 8}\} $   \\ 
$ 10 + 20\ F(z,\vec{b}) - 30\ G(z,\vec{b})$    &  $\ \{\mbox{\bf 10}\} $    \\ 
\end{tabular},\right.
\label{probalb}
\ee
where
$F(z,\vec{b}) = \exp(- \mbox{\large$\frac{9}{8}$}\ \sigma^{in}_{NN}\ T(z,\vec{b}))$
and
$G(z,\vec{b}) = \exp(- \mbox{\large$\frac{3}{4}$}\ \sigma^{in}_{NN}\ T(z,\vec{b}))$.
One notices several properties of the probabilities. First of all, the
limit $T \rightarrow 0$ implies $F,G \rightarrow 1$. This means that 
$P_1 \rightarrow 1$ and $P_8,P_{10} \rightarrow 0$. In other words the nucleon,
before entering the nucleus, is in singlet state as it should be. Moreover,
$P_1 + P_8 + P_{10} = 1$ for any $T$, therefore probability is always conserved.
Finally, if the nucleus is large enough
one has $T \gg 1$ and $F,G \ll 1$. This implies that if sufficient scatterings 
have taken place, the color probabilities reach the statistical limit, given by the 
first coefficients of eqs.$\,$(\ref{probalb}).
It turns out that soon after the nucleon has penetrated the nuclear 
profile, the statistical limit is essentially reached in which about $16/27 \simeq 2/3$
of the nucleons are in octet state while $10/27 \simeq 1/3$
are in decuplet. The amount of singlet is negligible.

With the calculated probabilities $P_n$ and the previously evaluated color cross
sections $\sigma^{abs}_{\Psi N_n}$, we can now construct the cross section
for $\Psi$ when it scatters with a nucleon, taking into account all its color 
states, and modify eq.$\,$(\ref{glauber}) by making the replacement
\be
\sigma^{abs}_{\Psi N}\ T_A(z_A,\vec{b}_A) \rightarrow \!\int_{- \infty}^{z_A} 
\!\!\!dz'_A\ \Sigma_B(\tilde{z}_B,\vec{b}_B)\ \rho_A(z'_A,\vec{b}_A)\,,
\ \ \mbox{where}\ \ \ \tilde{z}_B = z_B - (z'_A - z_A)\,\frac{1 + v_\Psi}{1 - v_\Psi}.
\ee
The position-dependent effective cross section
\be
\Sigma_B(\tilde{z}_B,\vec{b}_B) = \sum_{n} 
\ \sigma^{abs}_{\Psi\,N_n}\ P^B_{n}(\tilde{z}_B,\vec{b}_B)\,,
\label{effective}
\ee
represents absorption by nucleus $B$, taking into account the multiple scattering of 
its nucleons with nucleus $A$. Analogous expressions hold for $A$.

\begin{figure}[t]
\centerline{\psfig{figure=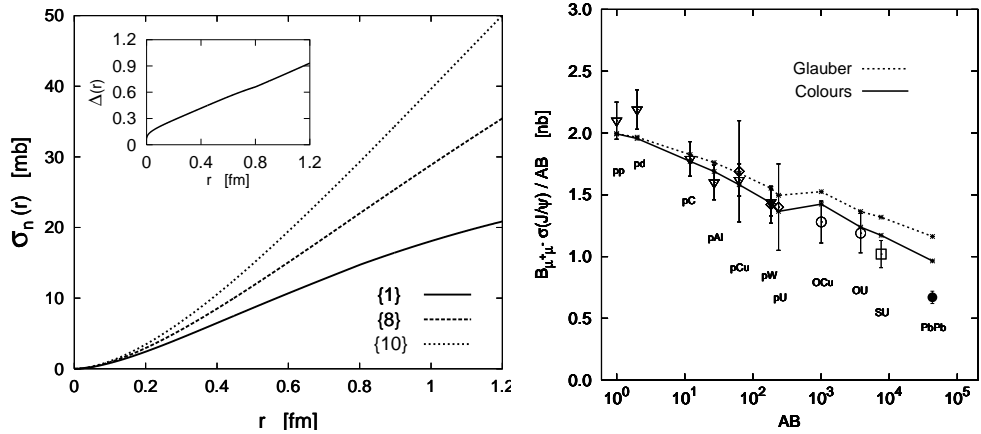,height=6cm,angle=-90}} 
\protect\caption{Left: Dipole cross section for different color states as a function
of the quark separation $r$. Box: The quantity $\Delta(r)$ which governs the 
dependence on color. Right: Calculations of total J/$\psi$ production cross sections
in $AB$ collisions and comparison with the NA38/50/51 data. The curves, drawn to guide
the eye, correspond to Glauber absorption (dashed) and to absorption by colored
nucleons (full).}
\label{total}
\end{figure}

The modified version of eq.$\,$(\ref{glauber}) can now be used to evaluate total 
cross sections, with
a further integration in impact parameter. We compute values to compare with 
available data measured by the NA38/50/51 collaboration at the SPS \cite{NA5097}.
We use realistic nuclear profiles and scale the center of mass energy according to
the parameterization $\sigma_\Psi^{NN} \sim (1 - m_{J/\psi}/\sqrt{s})^{12}$.
The results are compared with the data in Fig.~\ref{total}.
The dashed curve corresponds to a standard Glauber model calculation with the
singlet absorption cross section $\sigma^{abs}_{\Psi N_1} = 3.8$ mb, as
evaluated in Sect.~3, while the thick curve refers to the improved calculation 
which takes into account color excitations, studied in Sect.~4. In order to 
emphasize the difference between the two absorption mechanisms, we neglect 
here anti-shadowing effects.
For the $pA$ calculation we increased the value of $\sigma^{abs}_{\Psi N_1}$ by
35$\,\%$, therefore used $\sigma^{abs}_{\Psi N_8} = 5.1$ mb. This is the extreme
situation, when $\Psi$ always encounters color octet nucleons. For the $AB$ case
we used eq.$\,$(\ref{effective}).

Different from the Glauber model result, our improved one provides a better agreement 
with the data, except for the Pb+Pb point. In other words, the proposed absorption 
mechanism due to color dynamics is found to be an important contribution
to nuclear effects in $\Psi$ production. It must be accounted for, together with
other mechanisms of suppression  which we here put aside.

To summarize, we studied the dynamics underlying $\Psi$ suppression
due to nuclear effects, employing the idea that color exchange processes in multiple
$NN$ collisions may play a significant role.
We calculated inelastic cross sections for $\Psi$ scattering on
colored nucleons and found an increase of about $35\,\%$ when the nucleon is in a color
octet state as compared with the singlet case, while the enhancement factor for
the color decuplet is about $70\,\%$.

Making use of available calculations of $\Psi N$ cross sections, we
included formation time effects and obtained the effective value for  $\Psi$ 
absorption on singlet nucleons $\sigma^{abs}_{\Psi N_1} = 3.8$ mb. Scaling this
value with a $35\,\%$ increase we obtained $\sigma^{abs}_{\Psi N_8} = 5.1$ mb
and with a $70\,\%$ increase $\sigma^{abs}_{\Psi N_{10}} = 6.4$ mb.
We then performed a fit to available data on $\Psi$ production cross sections
in $pA$ collisions. The result of the fit, including gluon anti-shadowing 
corrections, is that the effective absorption cross section for $\Psi$ is
$\sigma^{eff}_{\Psi N} = 5.7 \pm 0.6$ mb, therefore suggesting that in $pA$ 
collisions, the produced charmonium interacts with color octet nucleons. The
conclusion is, though, rather weak because the experimental points are quite
scattered.

Total production cross sections for $\Psi$ were then calculated and compared in
Fig.~\ref{total} with available data for $AB$ collisions. Compared to the Glauber
model result, the agreement is improved and we conclude that effects due 
to color exchange processes are
substantial and provide the value $\sigma^{abs}_{\Psi N} = 5.5$ mb for the
effective absorption cross section on nucleons in $AB$ collisions (See 
eq.$\,$(\ref{effective})).

In conclusion, we have completed a study of nuclear effects in charmonium production
in $AB$ collisions, providing a better baseline for future calculations. These amount
to an improved treatment of gluon bremsstrahlung, including non-linearities arising
because of gluon cascades, to the calculation of effects caused by gluon energy loss
and to the evaluation of $E_\perp$-dependent cross sections. This will hopefully 
provide the starting point from which to address more precisely the nature of 
anomalous suppression in Pb+Pb collisions.

\end{document}